\documentclass[aps,prb,reprint,twocolumn,superscriptaddress,showpacs,amssymb]{revtex4-1}
\usepackage[english]{babel}
\usepackage{graphics}
\usepackage{graphicx}
\usepackage{epsfig}
\usepackage{amssymb}
\usepackage{dcolumn}
\usepackage{bm}
\usepackage{color}
\usepackage{natbib}
\usepackage{lineno}
\usepackage{sidecap}
\sidecaptionvpos{figure}{t}
\usepackage{verbatim}  
\usepackage{vector}  
\usepackage{wrapfig}
\usepackage{enumerate}
\usepackage{sidecap}
\usepackage{amsmath}
\usepackage{hyperref}

\begin{document}

\title{Effect of impurity substitution on band structure and mass renormalization of the correlated FeTe$_{0.5}$Se$_{0.5}$ superconductor}

\author{S.\ Thirupathaiah}
\email{t.setti@sscu.iisc.ernet.in}
\affiliation{Solid State and Structural Chemistry Unit, Indian Institute of Science, Bangalore, Karnataka, 560012, India.}
\author{J.\ Fink}
\affiliation{Leibniz Institut f\"ur Festk\"orper- und Werkstoffforschung IFW Dresden, D-01171 Dresden, Germany}
\author{P. K.\ Maheshwari}
\affiliation{CSIR-National Physical Laboratory, New Delhi 110012, India.}
\author{V. V. Ravi Kishore }
\affiliation{Solid State and Structural Chemistry Unit, Indian Institute of Science, Bangalore, Karnataka, 560012, India.}
\author{Z. -H. Liu}
\affiliation{Leibniz Institut f\"ur Festk\"orper- und Werkstoffforschung IFW Dresden, D-01171 Dresden, Germany}
\author{E. D. L. Rienks}
\affiliation{Leibniz Institut f\"ur Festk\"orper- und Werkstoffforschung IFW Dresden, D-01171 Dresden, Germany}
\author{B. B\"uchner}
\affiliation{Leibniz Institut f\"ur Festk\"orper- und Werkstoffforschung IFW Dresden, D-01171 Dresden, Germany}
\author{V. P. S. Awana}
\affiliation{CSIR-National Physical Laboratory, New Delhi 110012, India.}
\author{D. D. Sarma }
\affiliation{Solid State and Structural Chemistry Unit, Indian Institute of Science, Bangalore, Karnataka, 560012, India.}
\date{\today}

\begin{abstract}
Using angle-resolved photoemission spectroscopy (ARPES), we studied the effect of the impurity potential on the electronic structure of FeTe$_{0.5}$Se$_{0.5}$ superconductor by substituting 10\% of Ni for Fe which leads to an electron doping of the system.  We could resolve three hole pockets near the zone center and an electron pocket near the zone corner in the case of FeTe$_{0.5}$Se$_{0.5}$, whereas only two hole pockets near the zone center and an electron pocket near the zone corner are resolved in the case of Fe$_{0.9}$Ni$_{0.1}$Te$_{0.5}$Se$_{0.5}$, suggesting that the hole pocket having predominantly the $xy$ orbital character is very sensitive to the impurity scattering. Upon electron doping, the size of the hole pockets decrease and the size of the electron pockets increase as compared to the host compound. However, the observed changes in the size of the electron and hole pockets are not consistent with the rigid-band model. Moreover, the effective mass of the hole pockets is reduced near the zone center and of the electron pockets is increased near the zone corner in the doped Fe$_{0.9}$Ni$_{0.1}$Te$_{0.5}$Se$_{0.5}$ as compared to FeTe$_{0.5}$Se$_{0.5}$. We refer these observations to the changes of the spectral function due to the effect of the impurity potential of the dopants.
\end{abstract}
\pacs{ 74.70.Xa, 74.25.Jb, 79.60.-i, 71.20.-b }

\maketitle
\section{Introduction}
Most of the parent iron pnictides at ambient conditions,~\cite{Kamihara2008b, Rotter2008a, Jeevan2008, Tapp2008, Chu2009, Mizuguchi2009b} except LiFeAs ($T_c\approx18$ K)~\cite{Wang2008} and FeSe ($T_c\approx8$ K),~\cite{Hsu2008} are antiferromagnetic metals and they show superconductivity upon chemical doping or substitution.~\cite{Rotter2008,  Jeevan2008a, Tanatar2009, Massee2009b} Therefore, superconductivity in iron-based superconductors is very often induced by impurity substitution into the parent compound followed by suppressing the long range antiferromagnetic ordering.~\cite{Mazin2008, Kuroki2008} Hence, a good knowledge on how the impurity dopants suppress the antiferromagntic ordering and lead the system to superconductivity is very crucial to understand the mechanism of high-$T_c$ superconductivity in iron-based superconductors.


Theoretical works suggest that impurities act as scattering centers which lead to a broadening of the bands.~\cite{Zhang2009m, Wadati2010, Nakamura2011, Berlijn2012} These studies also point out that with the 3d transition element substitution for Fe, a part of the additional electrons from the transition metal remain localized at the constituents. However, there are several ARPES reports which demonstrate that the Co substitution for Fe donates the charge carriers to the host system according to a rigid band model.~\cite{Brouet2009,Thirupathaiah2010, Sekiba2009, Neupane2011, Dhaka2013} On the other hand, with the substitution of Ni and Cu for Fe the additional doping concentration is reduced, while for Zn the additional electrons are completely localized at the Zn ions.~\cite{Ideta2013, Nakamura2011} In contradiction to the ARPES studies, a report using x-ray absorption spectroscopy suggested that the Co substitution for Fe atom is nothing but a kind of isovalent substitution.~\cite{Merz2012} Recent consensus is that upon substitution of the 3d transition element for Fe (electron doping),  the volume of the electron and hole Fermi surfaces increase and decrease, respectively, qualitatively consistent with the rigid-band model, but the effective electron doping decreases in going from Co to Zn.~\cite{Ideta2013, Nakamura2011} Similarly,  the localization of the doped holes is noticed when Fe is replaced by Cr~\cite{Clancy2012} or Mn.~\cite{Texier2012}

Many ARPES studies, dealing with the effect of impurities on the electronic structure, are available on weakly correlated 122-type~\cite{Thirupathaiah2010, Sekiba2009, Neupane2011,Dhaka2013, Rienks2013} and 111-type~\cite{Thirupathaiah2012,Cui2013, Xing2014} iron pnictides. Intriguingly, till date, no ARPES study has been made on the more correlated charge doped 11-type iron chalcogenides which motivated us for the present study, though there are transport,~\cite{Williams2009b, Shipra2010a, Huang2010,Wang2015a} thermal~\cite{Thomas2009} and magnetic  measurements~\cite{Wen2013} reporting on this issue. These studies suggest that the transition metal substitution for Fe in FeTe$_{1-x}$Se$_x$ lead to a metal-insulator transition at high concentrations.

In this paper we report on the electronic band structure, the Fermi surface topology  and the spectral function analysis of the Fe$_{1-x}$Ni$_x$Te$_{0.5}$Se$_{0.5}$ compounds (x=0 and 0.1) using ARPES,  in order to understand the effect of Ni substitution on the electronic structure and the electronic correlations of FeTe$_{0.5}$Se$_{0.5}$ superconductor.  We could resolve three hole pockets near the zone center and an electron pocket near the zone corner in the case of FeTe$_{0.5}$Se$_{0.5}$, consistent with the band structure of the similar compounds,~\cite{Tamai2010,Yamasaki2010, Maletz2014, Ambolode2015, Zhang2015, Yi2015} whereas only two hole pockets are resolved near the center and an electron pocket is resolved near the corner of the Brillouin zone in the case of Fe$_{0.9}$Ni$_{0.1}$Te$_{0.5}$Se$_{0.5}$, suggesting that the hole pocket that has predominantly $xy$ orbital character is very sensitive to impurity scattering. We observe a decrease in the size of the hole pockets and an increase in the size of the electron pockets with Ni substitution, suggesting an effective electron doping. However, the observed change in the size of the electron and hole pockets is not consistent with the rigid band model. We further noticed that the mass renormalization is reduced near the zone center and increased near the zone corner in Fe$_{0.9}$Ni$_{0.1}$Te$_{0.5}$Se$_{0.5}$ as compared to FeTe$_{0.5}$Se$_{0.5}$.
\section{Experimental details}

Single crystals of Fe$_{1-x}$Ni$_x$Te$_{0.5}$Se$_{0.5}$ ($x$ = 0 and 0.1)  were grown at National Physical Laboratory in Delhi using self-flux. FeTe$_{0.5}$Se$_{0.5}$  shows superconducting transition at a $T_c$ $\approx$ 14 K, while Fe$_{0.9}$Ni$_{0.1}$Te$_{0.5}$Se$_{0.5}$ shows no superconductivity down to 2 K.~\cite{Maheshwari2015, Kumar2012}  ARPES measurements were carried out at BESSY II (Helmholtz Zentrum Berlin) synchrotron radiation facility at the UE112-PGM2b beam-line using the "1$^3$-ARPES"~\cite{Borisenko2012a,Borisenko2012b} and the "1$^2$-ARPES"  end stations equipped with SCIENTA R4000 analyzer and SCIENTA R8000 analyzer, respectively.

The measurements on FeTe$_{0.5}$Se$_{0.5}$ superconductors were measured at "1$^3$-ARPES" end station. The total energy resolution was set between 5 and 10 meV, depending on the applied photon energy.  Samples were cleaved $\textit{in situ}$ at a sample temperature lower than 20 K. The measurements were carried out at a sample temperature $T\approx$1 K. The measurements on Fe$_{0.9}$Ni$_{0.1}$Te$_{0.5}$Se$_{0.5}$ are measured at "1$^2$-ARPES" end station. The total energy resolution was set between 15 to 20 meV, depending on the applied photon energy.  Samples were cleaved $\textit{in situ}$ and measured at a sample temperature of $T\approx$50 K.


\section{Calculations}
To understand the experimental data we have performed a theoretical analysis of the electronic band structure of FeTe. The band structure calculations are done within the local density approximation (LDA) using the PAW pseudopotentials and the plane waves~\cite{Blochl_PAW, Kresse_PAW} as implemented in the Vienna Abinitio Simulation Package (VASP).~\cite{Kresse_Vasp1, Kresse_Vasp2, Kresse_Vasp3, Kresse_Vasp4} We used the experimental lattice constants, while the internal coordinates are freely relaxed. Monkhorst-Pack k-point mesh of 16$\times$16$\times$12 has been used for the Brillouin zone sampling. The plane wave cutoff energy was set at 400 eV.

\begin{figure*}
	\centering
		\includegraphics[width=0.95\textwidth]{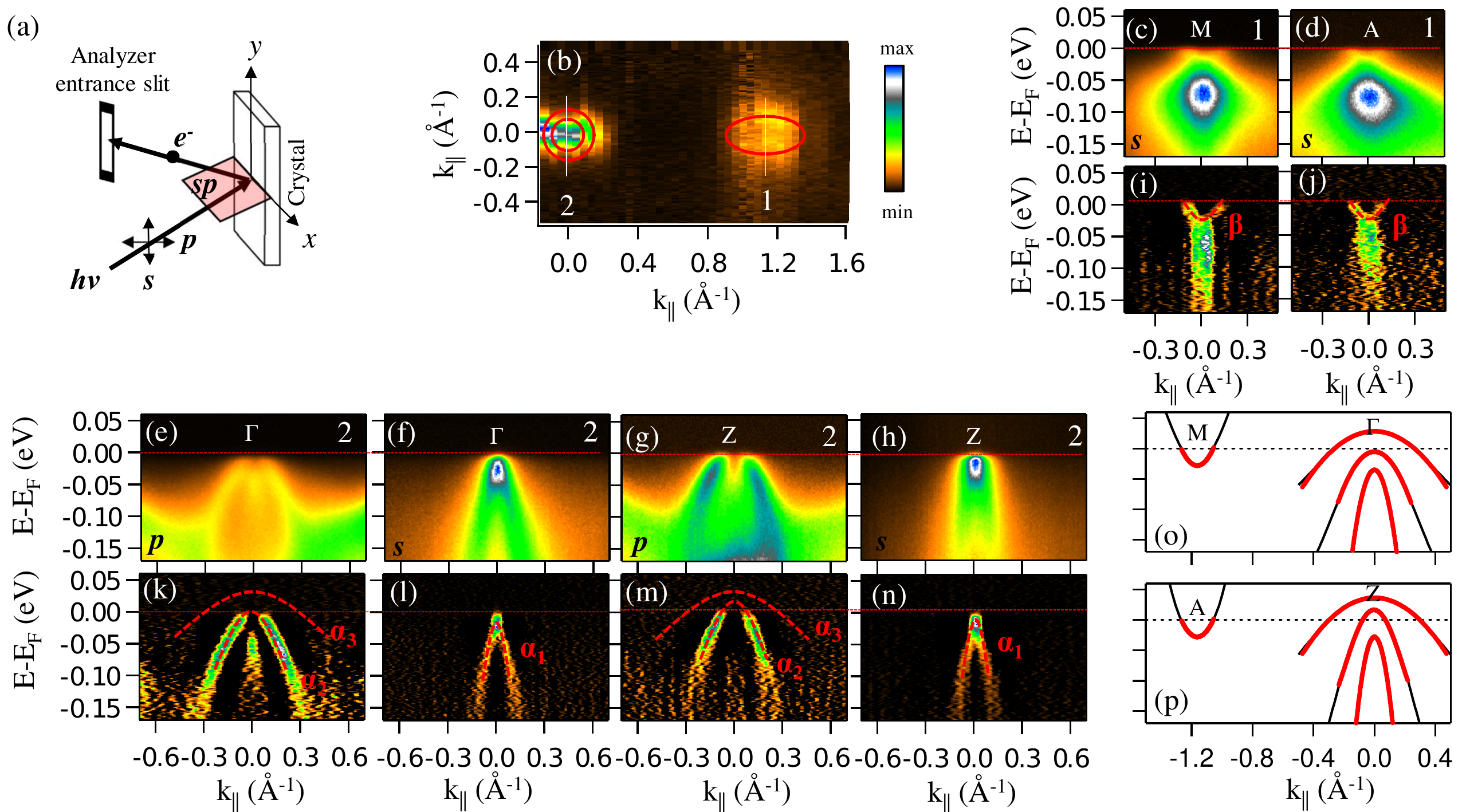}
	\caption{(Color online) ARPES data from FeTe$_{0.5}$Se$_{0.5}$. In (a) we schematically show our measuring geometry in which we define
$s$- and $p$-plane polarized lights with respect to the analyzer entrance slit. (b) is the Fermi surface map measured using $s$-polarized light with a photon energy h$\nu$=81 eV. (c) and (d) show the energy distribution maps (EDMs) taken  close to the high symmetry points $M$ and $A$ (see text), respectively, measured using $s$-polarized light along cut~1 as showed on the FS map. (e) and (f) show EDMs taken near $\Gamma$ measured using $p$- and $s$-polarized lights, respectively, along cut~2 as shown on the FS map. (g) and (h) are analogous data to (e) and (f), respectively, but taken near the $Z$-point. (i)-(n) are the second derivatives of (c)-(h). The red colored contours in (b) are guides to the eye schematically representing the Fermi sheets and the curves in (i)-(n) are guides to the eye schematically representing the band dispersions. In (o) and (p) we show the renormalized DFT-LDA band structure (black curves) overlaid on the experimental (red curves) band dispersions. We have normalized the EDMs by higher orders of the monochromator above the Fermi level.}
	\label{1}
\end{figure*}

\section{Results}
\begin{figure*}
	\centering
		\includegraphics[width=0.7\textwidth]{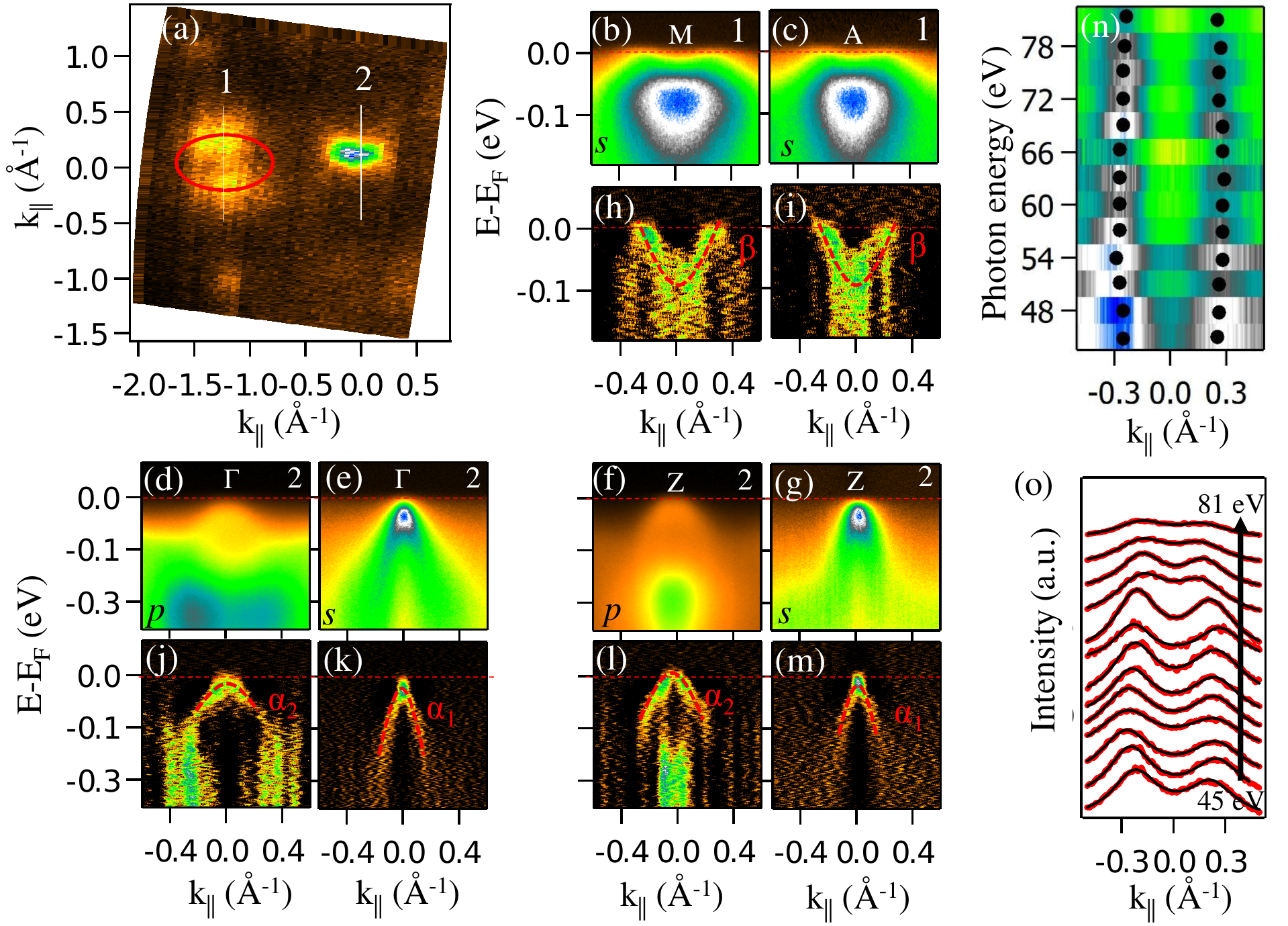}
	\caption{(Color online) ARPES spectra of Fe$_{0.9}$Ni$_{0.1}$Te$_{0.5}$Se$_{0.5}$. (a) is the Fermi surface map measured using $s$-polarized light with a photon energy h$\nu$=81 eV. (b) and (c) show the energy distribution maps (EDMs) taken close to the high symmetry points $M$ and $A$ (see the text), respectively, measured using $s$-polarized light along the cut~1 as shown on the FS map. (d) and (e) are EDMs taken near $\Gamma$ measured using $p$- and $s$-polarized lights, respectively along the cut~2 as shown on the FS map. (f) and (g) are analogous data to (d) and (e), respectively, but taken near the $Z$-point.  (h)-(m) are the second derivatives of (b)-(g). (n) is the Fermi surface map in the $k_y-k_z$ plane at the zone corner. The black circles are overlaid on the FS map representing the peak positions of the $\beta$ band. (o) shows a stack-plot of momentum distribution curves (MDCs) sampling different $k_z$ (red curves), together with results of a fit using a pair of Lorentzian functions. The red colored contour in (a) is guide to the eye schematically representing the Fermi sheet and the dashed curves in (h)-(m) are guides to the eye schematically representing the band dispersions. We have normalized the EDMs by higher orders monochromator above the Fermi level.}
	\label{2}
\end{figure*}

Figure~\ref{1} shows ARPES data from FeTe$_{0.5}$Se$_{0.5}$ superconductor. Fig.~\ref{1}(a) depicts our measuring geometry where we define the $s$- and $p$-plane polarized lights with respect to the analyzer entrance slit. Fig.~\ref{1}(b) is the Fermi surface map measured using $s$-polarized light with an excitation energy h$\nu$=81 eV. The data were recorded at a sample temperature of 1 K. To avoid influence of the SC gap, the FS map is extracted by integrating over an energy window of 10 meV centred at $E_F$, in which the hole pockets at the zone center and the electron pockets at the zone corner are seen.   Figs.~\ref{1}(c) and ~\ref{1}(d) show energy distribution maps (EDMs) taken along the cut~1 as shown on the FS map measured using the $s$-polarized light with photon energies 58 eV and 45 eV, respectively. According to the equation, $k_{\bot} = \sqrt{\frac{2m_e}{\hbar ^2} [E_{kin} cos^2\theta+V_0]}$  (the inner potential $V_0=15\pm3$ eV~\cite{Thirupathaiah2010}), near the zone corner 58 eV photon energy detects the bands at a $k_z=3.8~\pi/c$ and the 45 eV photon energy detects the bands at a $k_z=3.4~\pi/c$.  We assume that the used photon energies 58 eV and 45 eV are nearly close to the high symmetry points $M$ and $A$, respectively. Figs.~\ref{1}(e) and ~\ref{1}(f) are the EDMs measured using the $p$- and $s$-polarized lights with a photon energy of 58 eV, respectively along the cut~2 as showed on the FS map.  Figs.~\ref{1}(g) and ~\ref{1}(h) are the analogous data to \ref{1}(e) and \ref{1}(f), but measured with a photon energy of 45 eV.  Figs.~\ref{1}(i)-(n) are the second derivatives of ~\ref{1}(c)-(h), respectively. Using the photon energy 58 eV one could detect the bands at a $k_z=4.06~\pi/c$ (close to $\Gamma$) and with 45 eV one could detect the bands at a $k_z=3.6~\pi/c$ (close to $Z$) near the zone center. All the EDMs showed in Fig.~\ref{1} are recorded along the $\Gamma-M$ high symmetry line.
\begin{figure*}
	\centering
		\includegraphics[width=0.95\textwidth]{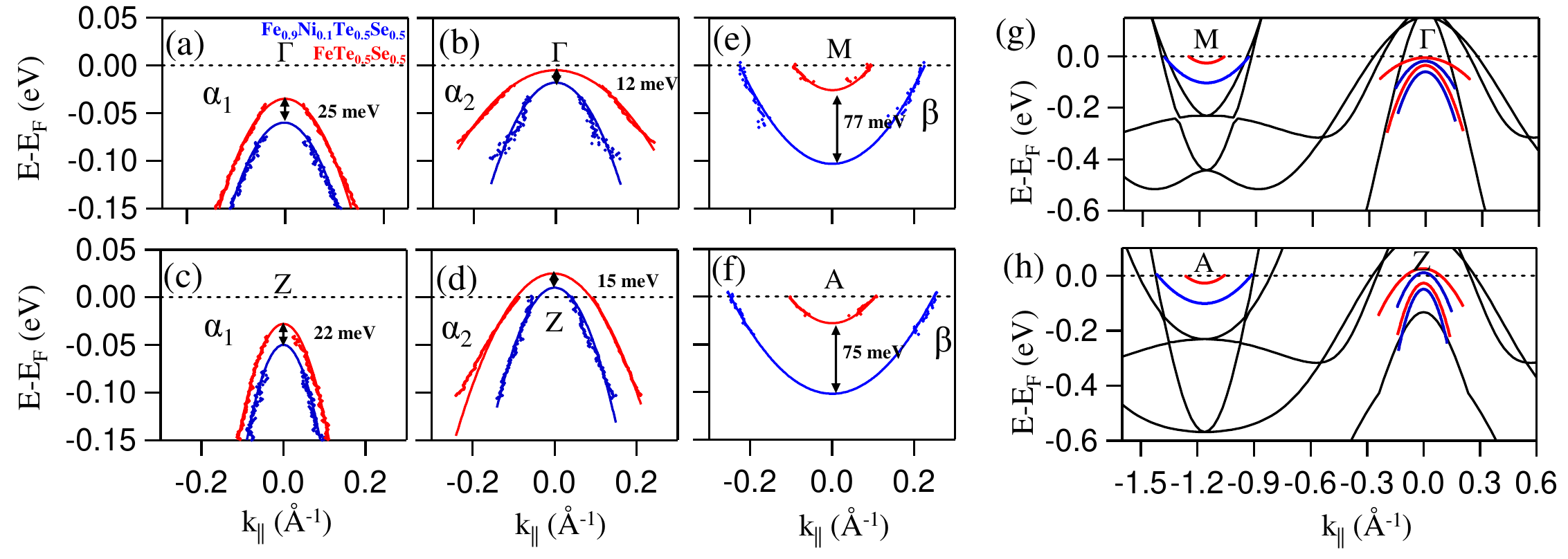}
		\caption{(Color online) In the figure the red and blue curves are the band dispersions from FeTe$_{0.5}$Se$_{0.5}$ and Fe$_{0.9}$Ni$_{0.1}$Te$_{0.5}$Se$_{0.5}$, respectively resulted from a fit to the MDCs using a pair of Lorentzian functions at $\Gamma$ for the band $\alpha_1$ (a) and for $\alpha_2$ (b). Similarly the band dispersions are shown at $Z$ for $\alpha_1$ (c) and for $\alpha_2$ (d). Panels (e) and (f) depict the band dispersions from $M$ and $A$, respectively.  (g) shows the DFT-LDA (black curves) band structure overlaid with the experimental band dispersions along the $\Gamma-M$ high symmetry line.  Similarly, (h) shows the band structure overlaid with the experimental band dispersions along the $Z-A$ high symmetry line. }
	\label{3}
\end{figure*}

From Figs.~\ref{1}(c) and \ref{1}(d) we could resolve an electronlike band $\beta$, crossing the Fermi level at a momentum vector $k_F=0.08\pm0.02~\AA^{-1}$ near the $M$-point, while near the $A$-point it crosses $E_F$ at a $k_F=0.1\pm0.02\AA^{-1}$. Thus, the band $\beta$ shows no $k_z$ dispersion within error bars in going from $M$ to $A$. In Figs.~\ref{1}(e)-(h) we could resolve three holelike bands, $\alpha_1$, $\alpha_2$ and $\alpha_3$, at the high symmetry points $\Gamma$ and $Z$. A Fermi vector $k_F=0.22\pm0.02~\AA^{-1}$ for $\alpha_3$ band is estimated by comparing the experimental data with DFT calculations  both at $\Gamma$ and $Z$. With the help of first principle calculations and polarization dependent selection rules, we ascribe $xy$ orbital character to the outermost hole pocket $\alpha_3$, while $xz/yz$ character to the hole pockets $\alpha_1$, $\alpha_2$ and the electron pocket $\beta$. Generally, the photoemission cross section is low for the inplane orbitals ($xy$), hence the band $\alpha_3$ is  scarcely resolved both at $\Gamma$ and $Z$ [see Figs.~\ref{1}(e) and (g)]. The band $\alpha_2$ disperses strongly towards $E_F$ but does not cross it, forming a van Hove singularity near the Fermi level at the $\Gamma$-point, consistent with the iron pnictide superconductors.~\cite{Liu2010b, Zabolotnyy2009a, Borisenko2010, Lubashevsky2012, Thirupathaiah2013} On the other hand, near the $Z$-point it crosses $E_F$ at $k_F=0.11\pm0.02~\AA^{-1}$ and thus the band $\alpha_2$ is showing a finite $k_z$ dispersion. The band $\alpha_1$ is always below the Fermi level at both $\Gamma$ and $Z$, hence does not contribute to the Fermi surface.

In order to match the DFT band structure with the experimental dispersions [see Figs.~\ref{1}(o) and (p)] the bands,  $\alpha_1$, $\alpha_2$ and $\alpha_3$, are renormalized by a factor of 2, 2.6 and 4.6 and shifted by -78 meV, -48 meV and -3 meV, respectively near the $\Gamma$-point. Similarly, near the $Z$-point, $\alpha_1$, $\alpha_2$ and $\alpha_3$, are renormalized by a factor of 0.65, 2 and 4.6 and shifted by +170 meV, -117 meV and +8 meV, respectively. The band $\beta$ is renormalized by a factor of 2.5 and 0.65 and shifted by +65 meV and +327 meV at $M$ and $A$, respectively. Here negative energies represent band shifting towards higher binding energies and positive energies represent band shifting towards lower binding energies.

Figure~\ref{2} shows ARPES data from an electron doped Fe$_{0.9}$Ni$_{0.1}$Te$_{0.5}$Se$_{0.5}$ compound. Fig.~\ref{2}(a) depicts the Fermi surface map measured using $s$-polarized light with an excitation energy h$\nu$=81 eV, extracted by integrating over an energy window of 10 meV centred at $E_F$, in which hole pockets at the zone center and electron pockets at the zone corner are seen. The data were recorded at a sample temperature of 50 K.   Figs.~\ref{2}(b) and ~\ref{2}(c) show the EDMs taken along the cut~1 as shown on the FS map,  measured using the $s$-polarized light at $M$  and $A$, respectively. Figs.~\ref{2}(d) and ~\ref{2}(e) show the EDMs taken along the cut~2 as shown on the FS map measured using the $p$- and $s$-polarized lights at $\Gamma$, respectively.  Figs.~\ref{2}(f) and ~\ref{2}(g) are the analogous data to \ref{2}(e) and \ref{2}(f), but measured at $Z$ point.  Figs.~\ref{2}(h)-(m) are the second derivatives of ~\ref{2}(b)-(g), respectively. All the EDMs shown in Fig.~\ref{2} are recorded along the $\Gamma-M$ high symmetry line.

From Figs.~\ref{2}(b) and \ref{2}(c) we could resolve an electronlike band $\beta$, crossing the Fermi level at a momentum vector $k_F=0.23\pm0.02~\AA^{-1}$ near the $M$-point, while near the $A$-point it crosses $E_F$ at $k_F=0.25\pm0.02\AA^{-1}$. Thus, the band $\beta$ shows no $k_z$ dispersion within error bars in going from $M$ to $A$. From Figs.~\ref{2}(d)-(g) we could resolve only two holelike bands, $\alpha_1$ and $\alpha_2$, at both the high symmetry points $\Gamma$ and $Z$. We again ascribe $xz/yz$ character to the bands $\alpha_1$, $\alpha_2$ and $\beta$. Here, the bands $\alpha_1$ and $\alpha_2$ do not cross the Fermi level near $\Gamma$, while only $\alpha_2$ crosses $E_F$ with a negligible Fermi vector $k_F=0.04\pm0.01\AA^{-1}$ near $Z$. Next we show ARPES measurements performed to reveal information on the $k_z$ dependent electronic structure near the zone corner. For this, photon energy dependent ARPES spectra were recorded along $k_{y}$, with photon energies ranging from h$\nu$=45 to 81 eV in steps of 3 eV. The data were recorded using the $s$-polarized light along the $\Gamma-M$ high symmetry line. Fig.~\ref{2}(n) depicts the Fermi surface map in the $k_y-k_z$ plane measured from Fe$_{0.9}$Ni$_{0.1}$Te$_{0.5}$Se$_{0.5}$, extracted by integrating over a window range of 15 meV centered at $E_F$.  Figure~\ref{2}(o) shows stack-plot of momentum distribution curves as a function of photon energy, fitted with a pair of Lorentzian functions. The peak positions of the $\beta$ band near the Fermi level extracted from the fits are shown by the black circles on the FS map [see Fig.~\ref{2}(n)], again suggesting that the electron pocket shows a weak $k_z$ dispersion.

In Fig.~\ref{3} the red and blue curves are the band dispersions from FeTe$_{0.5}$Se$_{0.5}$ and Fe$_{0.9}$Ni$_{0.1}$Te$_{0.5}$Se$_{0.5}$, respectively, resulted from a fit to the MDCs using a pair of Lorentzian functions at $\Gamma$ for the band $\alpha_1$ [Fig.~\ref{3}(a)] and for $\alpha_2$ [Fig.~\ref{3}(b)]. Similarly the band dispersions are shown at $Z$ for $\alpha_1$ [Fig.~\ref{3}(c)] and for $\alpha_2$ [Fig.~\ref{3}(d)]. Panels ~\ref{3}(e) and ~\ref{3}(f) depict the band dispersions from $M$ and $A$, respectively.  In Figs.~\ref{3}(a)-(f) the solid curves are parabolic fits to the band dispersions from which we could extract the effective masses ($m^*$). Fig.~\ref{3}(g) shows the DFT-LDA band structure (black curves) with overlaid experimental band dispersions along the $\Gamma-M$ high symmetry line.  Similarly, Fig.~\ref{3}(h) shows the DFT-LDA band structure with overlaid experimental band dispersions along the $Z-A$ high symmetry line.

\section{Discussions}
 We could resolve three hole pockets, $\alpha_1$, $\alpha_2$ and $\alpha_3$,  around $\Gamma$ and $Z$ and an electron pocket, $\beta$, around $M$ and $A$ from the FeTe$_{0.5}$Se$_{0.5}$ superconductor (see Fig.~\ref{1}). This observation is consistent with the ARPES reports on similar compounds.~\cite{Tamai2010,Yamasaki2010, Ambolode2015, Zhang2015, Yi2015} Effective mass enhancements are estimated for this compound upon employing  parabolic fits (see Fig.~\ref{3}) to the experimental band dispersions ($m^*$) and the DFT band structure ($m_b$). We have calculated a mass renormalization factor ($m^*/m_b$) for the bands $\alpha_1$, $\alpha_2$ and $\alpha_3$ of 1.77$\pm$0.04, 2.23$\pm$0.05 and 4.76$\pm$0.20, respectively near $\Gamma$ and 0.62$\pm$0.01, 1.74$\pm$0.22 and 4.76$\pm$0.20, respectively near $Z$. Similarly, the band $\beta$ near $M$ and $A$ shows a mass  renormalization factor of 1.78$\pm$0.16  and 0.66$\pm$0.03,  respectively. The mass renormalization values extracted from the parabola fits are very much close to the values extracted independently by scaling the DFT band structure, as discussed in the previous section. Here we can notice that the mass renormalization factors are relatively smaller at the $Z$-point compared to the $\Gamma$-point for the bands $\alpha_1$, $\alpha_2$ and $\beta$, which further suggests $k_z$ dependent correlations in these compounds. On the whole our findings on the band dependent mass renormalization factors ranging from 1.7-5 observed in the $\Gamma-M$ plane are in very good agreement with previous reports~\cite{Yamasaki2010,Aichhorn2010, Maletz2014, Ambolode2015, Zhang2015, Yi2015, Thirupathaiah2015}, while contradicting to the mass renormalization factor of 17 reported in Ref~\onlinecite{Tamai2010}. However, Ref.~\onlinecite{Fink2015} has reported a huge mass renormalization at the Lifshitz transition in the ferropnictides, but nevertheless at higher binding energies the mass renormalization is not far from the present values.


 Coming to the main results of this paper, we could resolve only two hole pockets near $\Gamma$ and $Z$ in the case of electron doped Fe$_{0.9}$Ni$_{0.1}$Te$_{0.5}$Se$_{0.5}$ compound. Recently, a similar observation has been made on a non-Fe-stoichiometric  Fe$_{1.068}$Te$_{0.54}$Se$_{0.36}$ compound in which case also only two hole pockets are resolved.~\cite{Thirupathaiah2015} Note here that the former compound is a non-superconductor,  while the latter one is a superconductor. However, from both compounds the third hole pocket ($\alpha_3$) mainly composed of the $xy$ orbital is unable to detect experimentally. From this, we can understand that $xy$ hole pocket is very sensitive to the system's stoichiometry and the impurity scattering. It could be due to the fact that $\alpha_3$ has low scattering cross section in ARPES measurements and has higher band renormalization compared to the other hole pockets which further broadens the spectral function, we are unable to detect this band from our ARPES studies when the impurities are added. However, theoretically we do not exclude its presence in Fe$_{0.9}$Ni$_{0.1}$Te$_{0.5}$Se$_{0.5}$.  Importantly  we noticed that, upon electron doping, in Fe$_{0.9}$Ni$_{0.1}$Te$_{0.5}$Se$_{0.5}$ the chemical potential shifts in accord with a non-rigid band model, unlike in 122-type and 111-type compounds where the impurity substitution shifts the bands in a rigid band model.~\cite{Ideta2013, Nakamura2011, Berlijn2012} It is important to note here that without inducing correlations to the system with impurity substitution, the charge doping lead to a rigid band model irrespective of the number of excess carriers contributing to the conduction band.~\cite{Ideta2013, Nakamura2011} Therefore, the observed non-rigid band type chemical potential shift in the studied compound is directly related to the induced correlation effects (discussed below) with the substituted impurity potential.~\cite{Berlijn2012}

 From Figs.\ref{3}(a)-(d) it is clear that the top of holelike bands near $\Gamma$ shifts towards higher binding energy almost by an average of 23 meV and near $Z$ it is of 13 meV upon the electron doping. On the other hand, near the zone corner [see Figs.~\ref{3}(e) and (f)] we noticed that the bottom of electronlike band shifts towards the higher binding energy almost by an average of 76 meV. These details indicate that the Ni doping is leading to an effective electron doping. Thus, the size of the hole pockets got shrunk and the electron pockets got increased, but unequally. That means the decrease in the hole pocket size is relatively low compared to the increase in the electron pocket. This observation is not supporting the experimental~\cite{Brouet2009,Thirupathaiah2010, Sekiba2009, Neupane2011, Dhaka2013, Merz2012, Cui2013, Xing2014} and theory~\cite{Ideta2013, Nakamura2011, Haverkort2011} reports on weakly correlated 122-type and 111-type compounds which suggest that in going from the substitution of Co to Ni to Cu to Zn in the place of Fe, the volume of the Fermi sheets increase and decrease which is qualitatively consistent with the rigid-band model.

 Interestingly, the mass renormalizations are reduced near the zone center and are increased near the zone corner upon the impurity substitution. More precisely, the effective mass ($m^*_{Ni}$) got reduced in Fe$_{0.9}$Ni$_{0.1}$Te$_{0.5}$Se$_{0.5}$ as compared to the FeTe$_{0.5}$Se$_{0.5}$ superconductor ($m^*$) by a factor ($m^*/m^*_{Ni}$) of  1.21$\pm$0.02 and 2.54$\pm$0.26 for $\alpha_1$ and $\alpha_2$, respectively near $\Gamma$, while 1.06$\pm$0.03 and 2.24$\pm$0.03 for $\alpha_1$ and $\alpha_2$, respectively near $Z$.  On the other hand, the effective mass ($m^*_{Ni}/m^*$) got increased by a factor of 1.4$\pm$0.1 and 1.5$\pm$0.1 at $M$ and $A$, respectively. Here we can notice that the estimated mass enhancements with Ni substitution are almost same within error bars both at $\Gamma$ and $Z$. This clearly demonstrates that Ni doping does not induce additional $k_z$ dependent correlations. In principle, the impurity substitution could affect the electronic structure of the host by three different ways: a) the crystal field splitting,  b) the average scattering potential which is generally determined by the onsite Coulomb energy (U) and c) the electron-impurity scattering. In the present case, Fe is partially replaced by Ni which has a similar ionic size as Fe. Therefore, the crystal field splitting in the electronic structure should be negligible.~\cite{Kumar2012} Similarly, the increase of onsite $U$ when going from  Fe to Ni is small~\cite{Sanchez-Barriga2012} which probably also indicates that this effect does not determine the observed effective mass changes. On the other hand, Refs.~\onlinecite{Berlijn2012} and ~\onlinecite{Haverkort2011} suggest an enhanced complex self-energy with the impurity substitution due to the electron-impurity scattering, which consequently enhances the real part of the self-energy and thus the effective mass. The increased effective mass for the electron pockets with Ni substitution is consistent with Ref.~\onlinecite{Haverkort2011}, however, the decreased effective mass for the hole pockets is in contrast and is further demonstrating the complexity in the microscopic understanding of the impurity substitution effect on the electronic structure of the iron-based compounds.~\cite{Berlijn2012} Moreover, there are several ARPES reports on this issue in the case of 122-type compounds which further demonstrate the complexity of understanding this problem.~\cite{Brouet2009, Xing2014, Dhaka2013} For instance, Refs.~\onlinecite{Brouet2009} and ~\onlinecite{Xing2014} show decreased electronic correlations, while Ref.~\onlinecite{Dhaka2013} shows unchanged electronic correlations upon the impurity substitution.

 Finally, in the FeTe$_{0.5}$Se$_{0.5}$ superconductor we could notice hole Fermi sheets at the $\Gamma$-point and electron Fermi sheets at the $M$-point, hence there the low energy interband scattering between the hole and the electron pockets (see Fig.~\ref{1}) is possible. On the other hand, in Fe$_{0.9}$Ni$_{0.1}$Te$_{0.5}$Se$_{0.5}$ near the $\Gamma$ point the hole pockets are completely filled by the electrons with Ni substitution and are stretched below the Fermi level (see Fig.~\ref{2}), thus the low energy interband scattering between hole and electron pockets which is believed to be important for high-$T_c$ superconductivity in these compounds~\cite{Mazin2008a,Kuroki2008, Graser2010} is totally suppressed. This could be a natural explanation on, why the Ni substitution does turn the compound from a superconductor to a non-superconductor. Before closing this section, we would like to explicitly mention that the temperature difference between the data of FeTe$_{0.5}$Se$_{0.5}$ and Fe$_{0.9}$Ni$_{0.1}$Te$_{0.5}$Se$_{0.5}$ does not affect the conclusions of this paper, because the estimated band dispersions (see Fig.~\ref{3}) based on analysis of the momentum distribution curves does not depend much on the measured sample temperature.

\section{Conclusions}
In conclusion, we have studied the effect of impurity potential on the electronic structure of FeTe$_{0.5}$Se$_{0.5}$ superconductor by substituting 10\% of Ni for Fe. We could resolve three hole pockets near the zone center and an electron pocket near the zone corner in the case of FeTe$_{0.5}$Se$_{0.5}$, whereas only two hole pockets near the center and an electron pocket near the corner of the Brillouin zone are resolved in the case of Fe$_{0.9}$Ni$_{0.1}$Te$_{0.5}$Se$_{0.5}$, suggesting that the third hole pocket having predominately $xy$ orbital character is very sensitive to the impurity scattering. We observe a decrease in the size of the hole pockets and an increase in the size of the electron pockets with Ni substitution, suggesting an effective electron doping. However, the change in the size of the electron and hole pockets is not consistent with the rigid band model. We further noticed that the effective mass of the hole pockets is reduced near the zone center and the effective mass of the electron pockets is increased near the zone corner in Fe$_{0.9}$Ni$_{0.1}$Te$_{0.5}$Se$_{0.5}$ when compared to FeTe$_{0.5}$Se$_{0.5}$. We suggest that the peculiarity of the non-rigid band changes of the chemical potential with Ni substitution is directly related to reduced correlations at the zone center and increased correlations at the zone corner. We could notice the interband scattering between the hole and electron Fermi sheets in the FeTe$_{0.5}$Se$_{0.5}$ superconductor, whereas in non-superconducting Fe$_{0.9}$Ni$_{0.1}$Te$_{0.5}$Se$_{0.5}$ compound the interband band scattering is suppressed as the hole pockets near the $\Gamma$-point are completely filled by the added electrons, suggesting that the Fermi surface topology is essential for high-$T_c$ superconductivity in these compounds.

\section{Acknowledgements}
T.S. acknowledges support by the Department of Science and Technology (DST) through INSPIRE-Faculty program (Grant number: IFA14 PH-86). T.S. acknowledges greatly the travel support given by IFW Dresden for part of the measurements. J.F. and B.B. acknowledge support by the German Research Foundation (DFG) through the priority program SPP1458. V.P.S.A and P.K.M acknowledge the financial support from the Govt. of India through the DAE-SRC outstanding researcher award scheme.
\bibliography{FeTe}

\end{document}